\documentclass[aps,superscriptaddress,reprint]{revtex4-2}

\usepackage[utf8]{inputenc}
\usepackage{xcolor}
\usepackage{pst-node}
\usepackage{graphicx}
\usepackage{amsmath, amssymb, amsthm}
\usepackage{mathtools}
\usepackage{hyperref}
\usepackage[shortlabels]{enumitem}
\usepackage{dsfont}
\usepackage{amsthm}
\usepackage{amssymb}
\usepackage{amstext}
\usepackage{braket}
\usepackage{amsfonts}
\usepackage{nicefrac}
\usepackage{extarrows} 
\usepackage{graphicx}
\usepackage{relsize}
\usepackage{soul}
\usepackage{xfrac} 

\newcommand{\skiptext}[1]{}

\begin{document}

\title{Mixed-precision ab initio tensor network state methods adapted for NVIDIA Blackwell technology via emulated FP64 arithmetic}

\author{Cole Brower}
\email{cbrower@nvidia.com}
\affiliation{%
NVIDIA, 2788 San Tomas Expressway, Santa Clara, CA 95051
}%

\author{Samuel Rodriguez Bernabeu}
\email{srodriguezbe@nvidia.com}
\affiliation{%
NVIDIA, 2788 San Tomas Expressway, Santa Clara, CA 95051
}%

\author{Jeff Hammond}
\email{jeffpapers@nvidia.com}
\affiliation{%
NVIDIA Helsinki Oy, Porkkalankatu 1, 00180 Helsinki
}%

\author{John Gunnels}
\email{jgunnels@nvidia.com}
\affiliation{%
NVIDIA, 2788 San Tomas Expressway, Santa Clara, CA 95051
}%

\author{Sotiris S. Xantheas}
\email{Sotiris.Xantheas@pnnl.gov}
\affiliation{%
Advanced Computing, Mathematics, and Data Division, Pacific Northwest National Laboratory, Richland, Washington 99354, USA}%
\affiliation{%
Department of Chemistry, University of Washington, Seattle, WA 98195, USA}

\author{Martin Ganahl}
\email{martin.ganahl@sandboxaq.com}
\affiliation{SandboxAQ, Palo Alto, California, USA}%

\author{Andor Menczer}
\email{menczer.andor@wigner.hu}
\affiliation{%
Strongly Correlated Systems Lend\"ulet Research Group,
Wigner Research Centre for Physics, H-1525, Budapest, Hungary
}%
\affiliation{%
E{\"o}tv{\"o}s Lor{\'a}nd University, P\'azm\'any P\'eter S\'et\'any 1/C, 1117 Budapest, Hungary
}%
\author{\"Ors Legeza}
\email{legeza.ors@wigner.hu}
\affiliation{%
Strongly Correlated Systems Lend\"ulet Research Group,
Wigner Research Centre for Physics, H-1525, Budapest, Hungary
}%
\affiliation{Dynaflex LTD, Zrínyi u 7, 1028 Budapest, Hungary}
\affiliation{
Institute for Advanced Study,Technical University of Munich, Germany, Lichtenbergstrasse 2a, 85748 Garching, Germany
}
\affiliation{Parmenides Stiftung, Hindenburgstr. 15, 82343, Pöcking, Germany}

\date{\today}

\begin{abstract} 
We report cutting-edge performance results via mixed-precision spin adapted \textit{ab initio} Density Matrix Renormalization Group (DMRG) electronic structure calculations utilizing the Ozaki
scheme for emulating FP64 arithmetic through the use of fixed-point compute resources.
By approximating the underlying matrix and tensor algebra with operations on a modest number of fixed-point representatives (``slices''), we demonstrate 
on smaller benchmark systems and for the active compounds of
the FeMoco and cytochrome P450 (CYP) enzymes with complete active space (CAS) sizes of up to 113 electrons in 76 orbitals [CAS(113, 76)] and 63 electrons in 58 orbitals [CAS(63, 58)], respectively, that the chemical accuracy can be reached with mixed-precision arithmetic.
We also show that, due to its variational nature, DMRG provides an ideal tool to benchmark accuracy domains, as well as the performance of new hardware developments and related numerical libraries.
Detailed numerical error analysis and performance assessment are also presented for subcomponents of the DMRG algebra by systematically interpolating between double- and pseudo-half-precision.
Our analyis represents the first quantum chemistry evaluation of FP64 emulation for correlated calculations capable of achieving chemical accuracy and emulation based on fixed-point arithmetic,
and it paves the way for the utilization of state-of-the-art Blackwell technology in tree-like tensor network state electronic structure calculations, opening new research directions in materials sciences and beyond. 

\end{abstract}

\maketitle

{\it Introduction:} 
Over the past decade, the rise of deep learning for artificial intelligence has sparked an unprecedented growing demand for computational resources for training and inference of machine learning (ML) models, further fueled by the advent and large-scale deployment of large language models (LLMs).
Graphics Processing Units (GPUs) are in many cases the key enabler for ML training and inference, and advances in hardware capabilities often
lead to new and more efficient training and inference paradigms ~\cite{gh200,mi300,googletpu, b100,groq}, with NVIDIA's latest Blackwell hardware generation as a prime example.
While AI model training is the prime application of GPUs today, scientists have been investigating how hardware accelerators may be useful in other areas of computational sciences, e.g. scientific computing for materials science and chemistry ~\cite{terachem, vasp, QE-2017, Ganahl-2019,lewis_tpu_dft, Ganahl-2023, Menczer-2023b,  Menczer-2024b,Xiang-2024,Agarawal-2025}. 
A key difference between ML model training and scientific computing is arithmetic precision requirements. ML training and inference are typically relatively insensitive to arithmetic precision reduction, a fact that hardware vendors exploit heavily to increase computational throughput for ML training. For example, NVIDIA's latest Blackwell generation is optimized for reduced precision arithmetic ~\cite{Dongarra-2024,SC-2024,b100}, leading to unprecedented computational throughput for ML training and inference. 
Scientific computing, however, places severe restrictions on arithmetic precision, with double-precision often being mandatory ~\cite{lewis_tpu_dft,Menczer-2024b}. 
Here, reduced precision arithmetic may severely impact many linear algebra subroutines, can cause instabilities and may lead to unreliable results.
Prominent examples of high-accuracy demands arise in electronic structure calculations ~\cite{Swart-2016,Khedkar-2021,Feldt-2022,Pantazis-2009,Sharma-2014c,Vinyard-2017,Lubitz-2019,Kerridge-2015,Spivak-2017,Gaggioli-2018,Trond-2011,Pyykk-2012,Reiher-2014a,Tecmer-2016,Wenjian-2020,Becke-2013,Benavides-2017}. Achieving quantitative agreement with experimental results for the electronic structure of a molecule or material requires solving a large, linear system of equations to extremely high accuracy. The typically achievable experimental accuracy in this context is $\sim$1.6mHa, also know as the chemical accuracy. 
In many cases, an accurate solution within this error margin requires prohibitive amounts of memory and computational resources, particularly for chemical compounds containing 
transition metal elements with many close-lying electronic states ~\cite{Bartlett-2005,Bartlett-2007,Lischka-2018,Evangelista-2018,Pulay-2011,Szalay-2012}.

A middle ground between these two extremes is the use of mixed-precision arithmetic, i.e. emulated high-precision arithmetic if required, and the use of single- or reduced-precision arithmetic if accuracy and stability permit it. Computational subroutines may flexibly switch between different arithmetic precision levels depending on user-specified speed vs. accuracy tradeoffs. Mixed-precision computing is posed to become increasingly important in the near future, where emulation of floating-point computation can play a major role,  allowing systems to flexibly perform not only emulated FP64 operations~\cite{Ootomo-2024a,Uchino-2025}, but a spectrum of FP operations, at precisions not natively supported by the hardware, offering time-to-solution and power efficiency gains~\cite{Dongarra-2024}.

In this context, tensor networks \cite{White-1993, White-1996, Fannes-1989, Ostlund-1995, Rommer-1997, Legeza-1996, Verstraete-2004a, Vidal-2007, Schollwock-2011,Verstraete-2023} are a computational paradigm that is particularly well-suited to benefit from recent advances in hardware design and acceleration \cite{Hager-2004,Stoudenmire-2013,Nemes-2014,Ganahl-2019, Milsted-2019,Brabec-2021,Zhai-2021,Gray-2021,Unfried-2023,Ganahl-2023,Menczer-2023a,Menczer-2023b,Menczer-2024a,Menczer-2024b,Xiang-2024,Menczer-2024c}. First risen to fame in the area of computational quantum many-body physics\cite{Wilson-1975, White-1993, White-1996, Fannes-1989, Ostlund-1995, Rommer-1997}, use cases have since expanded into quantum chemistry \cite{White-1999, Legeza-2003c, Chan-2003, Chan-2011, Szalay-2015a,Baiardi-2020}, machine learning \cite{stoudenmire_supervised, glasser_2019, glasser_2020,generative_mps, tn_anomaly_detection_1, tn_anomaly_detection_2, tomut2024compactifaiextremecompressionlarge, wang2025tensornetworksmeetneural, sengupta2022tensornetworksmachinelearning}, computer science \cite{oseledets_ttcross, dolgov_parallel}, and computational fluid dynamics \cite{gourianov_quantum-inspired_2022, peddinti_quantum-inspired_2024, gourianov_tensor_networks_2025, hulst_quantum-inspired_2025}.
By now, tensor networks have become one of the most promising and powerful approaches to tackle multi-reference systems in an approximate, yet highly accurate, fashion ~\cite{White-1992a,White-1999, Murg-2010a,Nakatani-2013,Kurashige-2014a, Kurashige-2014b,Murg-2015, Larsson-2022,Menczer-2024b,Baiardi-2020}.
Here, we demonstrate how our highly efficient, hardware-accelerated implementation of the density matrix renormalization group (DMRG) \cite{White-1993} method can be leveraged to perform highly accurate, {\it ab initio} quantum-chemical electronic structure calculations in mixed-precision arithmetic on NVIDIA's Blackwell ~\cite{blackwell} GPU architecture. 

Benchmark calculations of our highly-parallelized, GPU-accelerated and SU(2)-aware implementation of the DMRG algorithm~\cite{Menczer-2023a,Menczer-2023b,Menczer-2023c,Menczer-2024a,Menczer-2024b,Menczer-2024c} are presented on an NVIDIA DGX B200 GPU supercomputer
~\cite{b100} via emulated FP64 arithmetic~\cite{Ootomo-2024a,Dongarra-2024,SC-2024}. 
This represents the first quantum chemistry evaluation of FP64 emulation for correlated calculations capable
of achieving chemical accuracy and emulation based on fixed-point arithmetic; previously, FP64 emulation was evaluated for traditional mean-field (DFT) methods using FP16 precision~\cite{Dawson:emulation-DFT:2024}.
Our work forms a major milestone in validating this novel approach in electronic structure calculations and also paves the way for applications via state-of-the-art Blackwell technology-based hardware architectures~\cite{b100}.

{\it Theory of TNS/DMRG regarding error analysis:} 
The DMRG is a variational optimization method for finding the ground state of a model Hamiltonian $H$ over the space of so-called matrix product states (MPS) ansatz wave functions ~\cite{Schollwock-2011}. Given the description of a quantum chemical system~\cite{White-1999} in terms of $N$ spinful orbitals $\ket{i_n} = \{\ket{0}, \ket{\uparrow}, \ket{\downarrow}, \ket{\uparrow\downarrow}\}$, its quantum mechanical wave function can be written as
\begin{equation}
  \ket{\Psi_{MPS}}  = \sum_{\{i_k\}} \sum_{\{\alpha_p\}}[A_1]_{1\alpha_1}^{i_1} [A_2]_{\alpha_1\alpha_2}^{i_2} \dots [A_{N}]_{\alpha_{N-1}1}^{i_{N}} \ket{i_1\dots i_k}
\label{eq:mps}
\end{equation}
where $A^{i_n}_{\alpha_{n-1}\alpha_n}$ are order-3 tensors of dimension $(D_{n-1},4,D_n)$ except
for the first and the last orbitals where order-2 tensors appear. 
The numerical accuracy of the ansatz is determined by the ranks of the matrices, $D$, also known as the bond dimension, with higher ranks corresponding to higher accuracy. The exact solution is recovered at $D_n\sim4^n$ for $1\leq n\leq N/2$ ($D_n\sim4^{N-n}$ for $N/2\leq n\leq N$).
In practice, the full ranks are truncated to achieve an approximate solution. The memory and compute requirements 
scale as $O(N^2D^2)$ and $O(N^4D^3)$, respectively.
In this work, bond dimensions, $D$, are reported as SU(2) multiplets~\cite{Mcculloch-2002,Toth-2008,Menczer-2023b}.
The DMRG method performs sequential updates one tensor at a time, while keeping all 
other tensors fixed, such that the expected energy $\langle \Psi_{MPS}|H|\Psi_{MPS}\rangle$ is iteratively lowered.

The first key subroutine is simply binary tensor contraction via matrix multiplications, which is the basic workhorse for any tensor network algorithm.
The second subroutine carries out the update for tensor $[A_n]^{i_n}_{\alpha_{n-1}\alpha_n}$. The update is obtained from an iterative diagonalization of an effective Hamiltonian matrix in a truncated Hilbert space of dimension proportional to $D^2$, using a Krylov-based eigensolver (Davidson or L\'anczos method). The accuracy and stability of this solver is typically quite sensitive to arithmetic precision errors. 
The third subroutine is used to shift the optimization from $[A_n]^{i_n}_{\alpha_{n-1}\alpha_n}$ to $[A_{n+1}]^{i_{n+1}}_{\alpha_{n}\alpha_{n+1}}$, and employs a singular value decomposition (SVD)
on the optimized joined tensor $[A_{(n, n+1),opt}]^{i_n i_{n+1}}_{\alpha_{n-1}\alpha_{n+1}}$ \cite{Schollwock-2011}.
The accuracy of these three main algorithmic steps influences the overall error of the DMRG algorithm in a complex way.
Our implementation allows us to switch between different CPU- and GPU-based implementations for each of these steps, providing us with an ideal framework to test numerical libraries
that implement these operations.
The fact that DMRG is a variational method, 
i.e., that the true ground state energy is strictly approached from above, with an error that is determined by the size of the bond dimension $D$ 
~\cite{White-1992b,Legeza-1996,Legeza-2003a}, can be utilized to create a complex test bed to test how arithmetic precision errors affect accuracy and stability of a) the Krylov solver as a function of the residual error $\varepsilon$ thresholds used therein \cite{Noack-2005,Schollwock-2011} and 
b) the SVD truncation.
Therefore, from a technical point of view, DMRG provides an ideal tool to validate and benchmark recent hardware developments and numerical libraries by adjusting parameters,
even further approximating FP64-precision,
so that the final accuracy can be controlled rigorously and performance can be monitored for a broad range of error margins.

{\it Theory of emulated FP64 arithmetic:}
In the following we provide a short description of the key ideas of our FP64 emulation strategy, and refer the reader to the literature for more details ~\cite{Ootomo-2024a,Uchino-2025}.
Considering the multiplication of two matrices ${\bf C} ={\bf A B}$, the key strategy is as follows: 1) convert floating point values in the matrices to a fixed-point format with  
exponents shared across each row of $\bf A$ and each column of $\bf B$,
2) decompose the matrices ${\bf A, B}$ into "slices" ${\bf A}^i, {\bf B}^j$, $i,j\in\{1,\dots ,S\}$ of lower precision, 3) perform matrix multiplications in lower precision for all pairs $(i,j)$ independently, and 4) accumulate the products at high precision into the final result ${\bf C} = \sum_{ij}{\bf A}^i {\bf B}^j$.
The decomposition into lower-precision matrices can be done in several different ways, as explained in e.g. \cite{Uchino-2025}, but all approaches follow similar strategies.
For example, the multiplication of two fixed-point numbers, using seven INT8 slices (to hold 63 mantissa bits) 
to represent each requires 49 (7x7) element-wise multiply-adds (aggregated using a higher precision data type, e.g. INT32). 
Further efficiencies can be gained by e.g. ignoring the less significant results in the lower-triangular portion of the 7x7 grid above. 
After the results are aggregated, the individual contributions are converted back 
into FP64. 

{\it Numerical procedure:}
Our numerical analysis will be presented for various strongly-correlated molecules and chemical clusters, i.e., multi-reference problems. DMRG simulations have been performed on GPU accelerated NVIDIA DGX H100 and DGX B200 single nodes using both traditional FP64 double-precision as well as emulated FP64 arithmetic.
The simple formula that connects number of slices, $S$, to mantissa bit counts is $S=\rm{ceildiv}(\rm{mantissa bits+1},8)$, 
which gives mantissa bit setting 15, 23, 31, 39, 47, 55 when approximating double-precision in emulated mode for $S=2,3,4,5,6,7$. The $+1$ term accounts for the sign bit.

NVIDIA's pre-release cuBLAS library offers various interfaces to test and use emulation with environment variables or corresponding APIs.  These environment variables enable/disable emulation, specify the number of mantissa bits
or allow the system to dynamically determine this value, and enable eager/performant emulation strategies.
With NVIDIA's pre-release cuBLAS library, when no environment variables are set, one will still see native FP64.  When emulation is enabled but no mantissa bit count is set, the library determines how many extra bits are needed to maintain the accuracy of native FP64.
While emulation is powerful for compute-bound matrix-multiplications, it is often slower for memory-bound matrix multiplications. Eager mode attempts to use emulation irrespective of the performance characteristics of the matrix multiplication problem, whereas performant mode enables a layer of heuristics to choose to use emulation when it would provide a performance advantage.
In order to study the numerical properties of mixed precision floating point emulation in DMRG
we have used eager mode and set $\varepsilon=10^{-5}$ 
in the rest of the paper unless otherwise specified.
Moreover, this pre-release library, via performant mode, the default, has an early version of the heuristics that invoke emulation only when it makes sense to do so from a performance perspective. A subsequent publication will include a performance study using an improved cuBLAS library and performant mode. Finally, we remark that results were obtained using a pre-release cuBLAS binary and the data is subject to change upon official release of the library.

{\it Numerical benchmark:} The first problem we consider is the F$_2$ molecule on a CAS(18,18) model space~\cite{Legeza-2003a}, for which the exact full-CI reference energy is also available from exact diagonalization in  
FP64 precision.
In this work all energies will be given in Hartrees (a.u.).
\begin{figure}
    \centering        \includegraphics[width=0.48\textwidth]{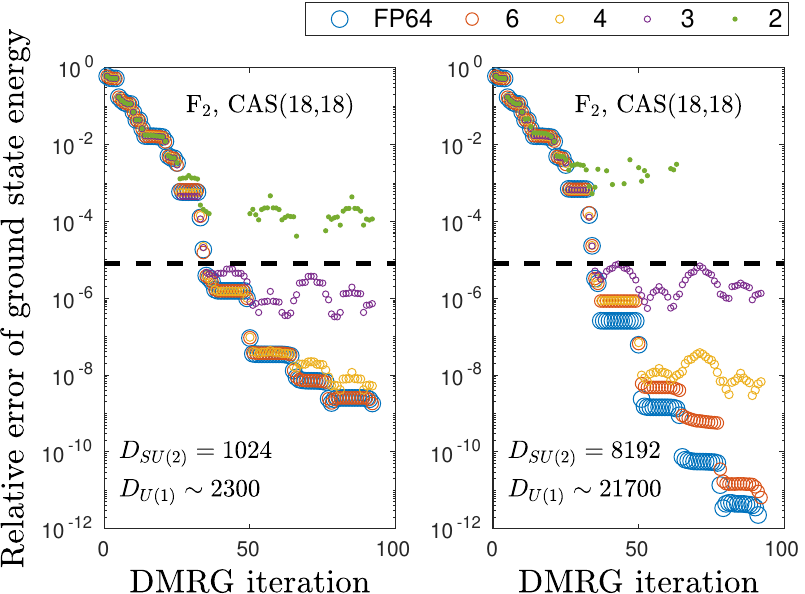}
    \caption{Relative error of the ground state energy as a function of DMRG iteration steps for the F$_2$ molecule in a CAS(18,18) model space using $D_{SU(2)}=1024$ (left panel) and $D_{SU(2)}=8192$ (right panel) SU(2) multiplets for the non-emulated native FP64 limit and for various number of INT8 slices, $S\in\{6,4,3,2$\} obtained on a DGX B200 system.
    The dashed line stands for the relative error of chemical accuracy.}
    \label{fig:egs_f2_m1024}
\end{figure}
In Fig.~\ref{fig:egs_f2_m1024} we show the relative error of the ground state energy as a function of DMRG iteration steps using $D_{SU(2)}=1024$ (left panel) and $D_{SU(2)}=8192$ (right panel) SU(2) multiplets for the non-emulated FP64 limit and for various number of INT8 slices, $S\in\{6,4,3,2\}$. 
We observe for both bond dimensions, $D$, 
that the FP64 reference limit is reproduced at $S=6$ slices, and a systematically increasing relative error with decreasing values of $S$. 
We note that $S=3$ slices are just sufficient to reach chemical accuracy for this system, but in general $S>3$ is required to obtain more reliable accuracy. Note that we only observe the expected increase in accuracy with increasing $D$ for $S\ge 4$. For $S=2$, i.e. half-like precision, we observe an error significantly above chemical accuracy, unstable DMRG iterations, and violations of the variational principle (i.e. energies below the exact solution, not shown in Fig.~\ref{fig:egs_f2_m1024}).

\begin{figure}
    \centering        \includegraphics[width=0.48\textwidth]{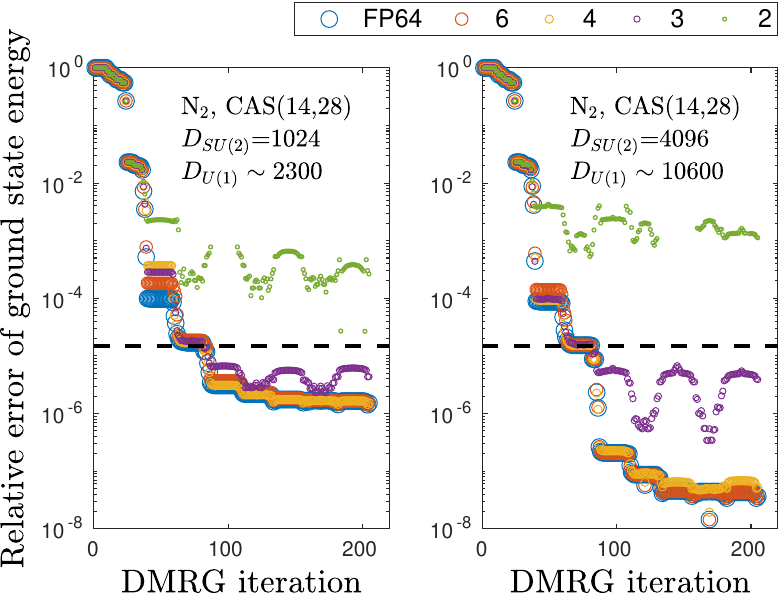}
    \caption{Similar to Fig.~\ref{fig:egs_f2_m1024} but for the nitrogen dimer at its equilibrium geometry in a CAS(14,28) model space using $D_{SU(2)}=1024$ (left panel) and $D_{SU(2)}=4096$ (right panel) SU(2) multiplets obtained on a DGX B200 system.
    }
    \label{fig:egs_n2_m1024}
\end{figure}

In Fig.\ref{fig:egs_n2_m1024} we repeat the same analysis for the more challenging case of the nitrogen dimer in the cc-pVDZ basis at equilibrium bond length, $r=2.118$a$_0$, and CAS(14,28), which has been subject of several DMRG benchmark calculations~\cite{Dunning-1989,Chan-2004b,Faulstich-2019b,Boguslawski-2013,Mate-2022,Menczer-2023b}. 
We find similar convergence profiles as summarized in Fig.~\ref{fig:egs_n2_m1024} 
The reference energy 
was determined through a highly accurate coupled-cluster, CCSDTQPH calculation~\cite{Chan-2004b} and validated by large-scale DMRG simulations up to six decimal digits~\cite{Chan-2004b,Menczer-2023b}. 
For both $D_{SU(2)}=1024$ and $D_{SU(2)}=4096$ (left and right panel, respectively) all emulated FP64 energies  with $S\ge3$ converged and reproduced the reference FP64 DMRG data within chemical accuracy, while calculations for $S=2$ again show poor accuracy, convergence issues, and non-variational energies.
Note that the expected  increase in accuracy with increasing bond dimension
is reproduced only at $S=4,6$ slices. 
In order to study how the eigenvalue spectrum affects numerical stability, we have repeated the same analysis, but for stretched geometries, i.e. for bond distances $r=3.600$a$_0$ and $4.200$a$_0$, where the multi-reference character is becoming more pronounced
~\cite{Boguslawski-2013,Faulstich-2019b,Mate-2022}. 
In general, we again obtain very similar and stable convergence profiles up to the error margins discussed above for $S>2$, and unstable simulations for $S=2$. 

\begin{figure}
    \centering        \includegraphics[width=0.48\textwidth]{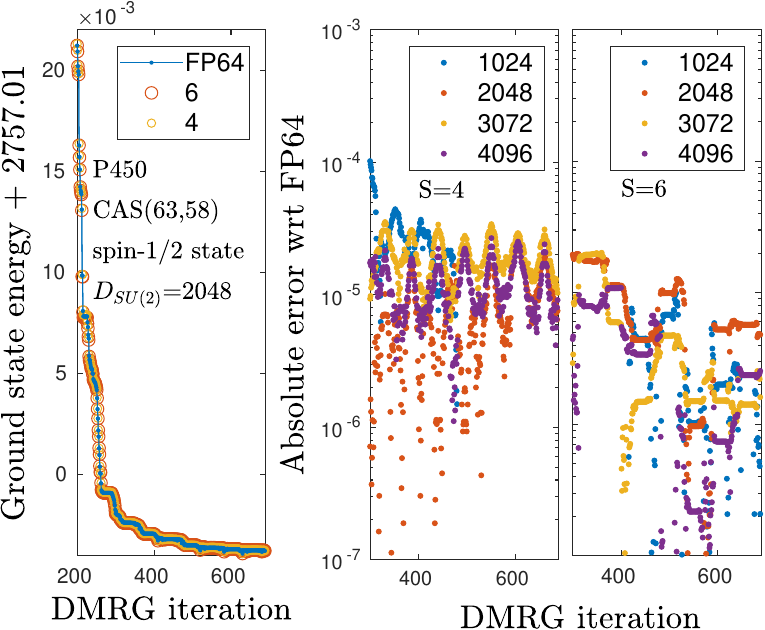}
    \caption{ 
    Shifted ground state energy for the spin-1/2 doublet state of the cytochrome P450 (CYP) enzymes with CAS(63,58) model space as a function of DMRG iteration
    using $D_{SU(2)}=2048$
    SU(2) multiplets and $S\in\{4,6\}$ slices (left panel) and the absolute error measured with respect to the native FP64 data sets for $S=4$ (central panel) and $S=6$ (right panel) slices for various $D$ values obtained on a DGX B200 system.
    }
    \label{fig:egs_oxoX_m1024}
\end{figure}

Next, we targeted a very complex chemical system, the cytochrome P450 (CYP) enzymes, using a CAS(63,58) model space introduced recently by the Google Research team~\cite{Goings-2022}. This system has also been in the focus of our research, demonstrating previously that our hybrid CPU-GPU DMRG code can reach 0.25 PFLOPS performance on a single DGX H100 node~\cite{Menczer-2024b}. Therefore, highly accurate DMRG reference data are available for the different spin states. In Fig.~\ref{fig:egs_oxoX_m1024} we present the obtained convergence profiles for the spin-1/2 doublet ground state with $D_{SU(2)}=2048$,
using both native FP64 and $S\in\{4,6\}$ slices (left panel), while the absolute error with respect to the native FP64 simulation is presented in the central and right panels for various bond dimensions.
It is clearly visible that stable convergence can be reached with $S\ge4$ number of slices. The absolute error for $S=6$ upon convergence is already in the range of $10^{-5}$ for $D_{SU(2)}=1024$, which drops further down to $10^{-6}-10^{-7}$ with increasing bond dimension. 
Similar results have been obtained for the spin-3/2 and spin-5/2 excited states. For $S=4$ a slightly larger error is obtained,
while for $S=3$ the absolute error was even larger than the chemical accuracy for all $D\ge 1024$ values.
Therefore,
using $S\in\{4,6\}$ slices we managed to reproduce our earlier results ~\cite{Menczer-2024b} for the spin gaps  within microHartree accuracy via mixed-precision arithmetic on state-of-the-art Blackwell hardware.  

Finally, for the FeMoco on CAS(54,54)~\cite{Reiher-2017,Kai-2020} and CAS(113,76) model spaces ~\cite{Li-2019,Brabec-2021,Menczer-2023b,Xiang-2024} we have found similar error profiles using $S\in\{4,6\}$ slices as discussed in
Fig.~\ref{fig:egs_oxoX_m1024}. In practice, the optimal number of slices are set automatically by default (performant mode) based on the most recent version of the pre-release cuBLAS library, which returned even slightly more accurate energy values. 
This let us conclude again that in electronic structure calculations, utilization of a
limited number of slices $S\in\{4,6\}$ is adequate 
to reach chemical accuracy.

\begin{figure}
    \centering        \includegraphics[width=0.48\textwidth]{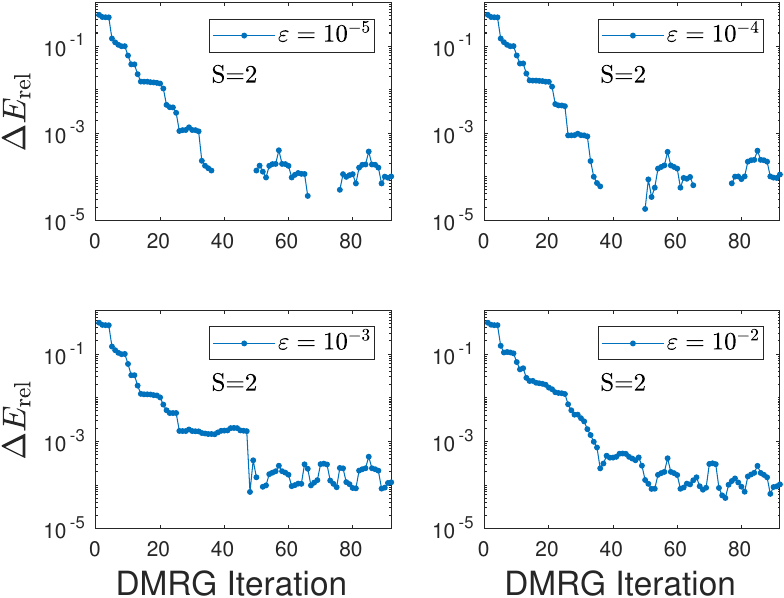}
    \caption{Relative error of the ground state energy, $\Delta E_{\rm rel}$, as a function of DMRG iteration steps for the F$_2$ molecule in a CAS(18,18) model space using bond dimension $D_{SU(2)}=1024$ and $S=2$ slices
    for various pre-set residual error threshold values, $\varepsilon$, employed in the L\'anczos diagonalization method. Note the missing data points due to non-variational eigenvalues.}
    \label{fig:egs_f2_m1024_lanczos}
\end{figure}

{\it Numerical stability:} To gather more details about the convergence of DMRG, its main algorithmic parts can be analyzed independently by switching between a CPU-based implementation or a CUDA version using double-precision or emulated FP64.
First, we found that 
the number of L\'anczos iteration steps for the diagonalization of the effective Hamiltonian
(for more detailed terminology see Refs.~\cite{Schollwock-2005,Szalay-2015a,Menczer-2023a,Menczer-2023b})
is almost the same for $S\ge3$ slices, in agreement with FP64 reference data. Consequently, the L\'anczos method is not sensitive to the enforced approximations used via the DGEMM operations~\cite{fp64-like}
and a stable convergence can be obtained. In contrast to this, for $S=2$ slices 
non-variational solutions have been returned for several DMRG iteration steps. This led to a complete failure of DMRG as shown in Figs.~\ref{fig:egs_f2_m1024}
and ~\ref{fig:egs_n2_m1024}. Reducing the residual error
$\varepsilon$ to $10^{-4},\ldots,10^{-2}$, the number of non-variational eigenvalues, however, disappeared leading to an oscillating curve in relative energy in the range of $10^{-4}$ as in Fig.~\ref{fig:egs_f2_m1024}, but without ``missing" data points (for more details, see Fig.~\ref{fig:egs_f2_m1024_lanczos} in the Supporting Information). We remark that we found the Davidson method less stable using emulated FP64 arithmetic with a reduced slice count, reflected by the slightly increased number of iterations during the iterative diagonalization procedure.    

\begin{figure}[htb]
    \centering        \includegraphics[width=0.48\textwidth]{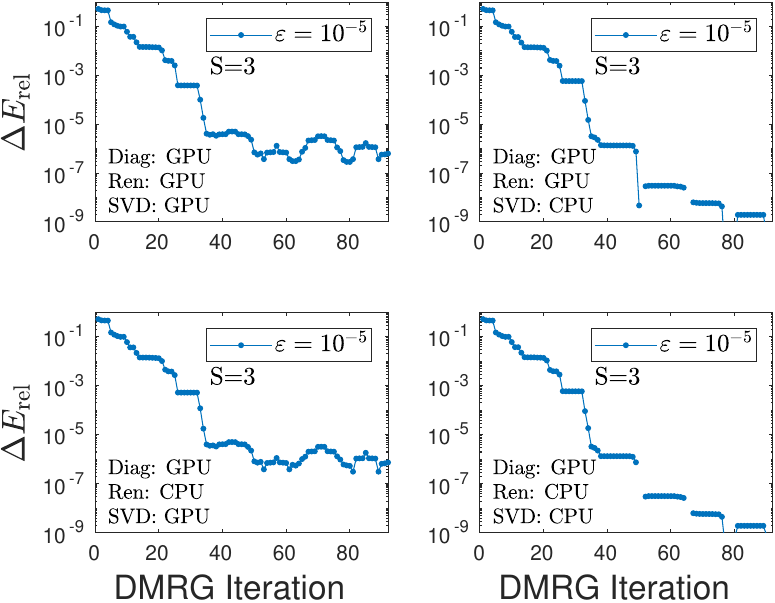}
    \caption{Relative error of the ground state energy, $\Delta E_{\rm rel}$ as a function of DMRG iteration steps for the F$_2$ molecule in a CAS(18,18) model space using bond dimension $D_{SU(2)}=1024$, $S=3$ slices,
    setting the residual error threshold in the L\'anczos diagonalization method to $\varepsilon=10^{-5}$ and employing various CPU and GPU implementations for the renormalization (network contraction) and singular value decomposition (SVD).}
    \label{fig:egs_f2_m1024_ren_svd_s3}
\end{figure}

\begin{figure}
    \centering        \includegraphics[width=0.48\textwidth]{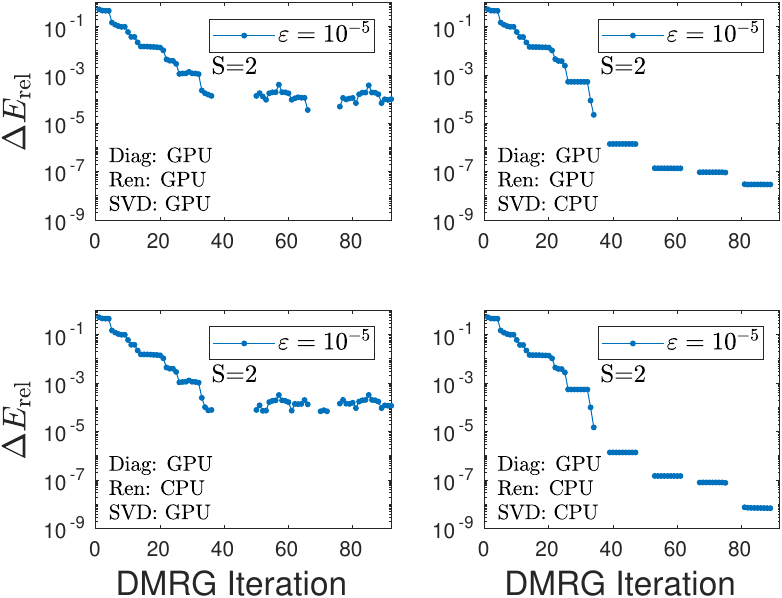}
    \caption{
    Similar to Fig~\ref{fig:egs_f2_m1024_ren_svd_s3} but for S=2 slices.
    }
    \label{fig:egs_f2_m1024_ren_svd}
\end{figure}

Next, we analyzed the effect of employing a CPU, via native FP64, or GPU, via emulated FP64, implementation of the network contraction, i.e. renormalization step, which is also based on DGEMM operations. In general, we obtained similar and stable convergence profiles for $S\ge3$ slices regardless of whether we used a CPU or GPU version.
Finally, we studied the effect of the eigenvalue solver offered by cuSOLVER to diagonalize the reduced density matrix and truncate Schmidt-spectrum accordingly. By employing the CPU-based Intel MKL LAPACK  function ({\it LAPACKE\_dsyevd}) via native FP64 or the 
GPU implementation ({\it cusolverDnXsyevd}) and carrying out the related linear algebra via reduced-precision emulated FP64-like arithmetic we have found 
significant effects on the convergence of the DMRG method for small $S\in\{2,3\}$ slices, and the error accumulated via SVD determines the overall convergence
(for more details see Figs.~\ref{fig:egs_f2_m1024_ren_svd_s3} and ~\ref{fig:egs_f2_m1024_ren_svd} in the Supporting Information). 
A more rigorous error analysis, 
also employing the dynamic block state selection (DBSS) approach~\cite{Legeza-2003a,Legeza-2003b},
will be part of a subsequent publication.

\begin{figure}
    \centering        \includegraphics[width=0.48\textwidth]{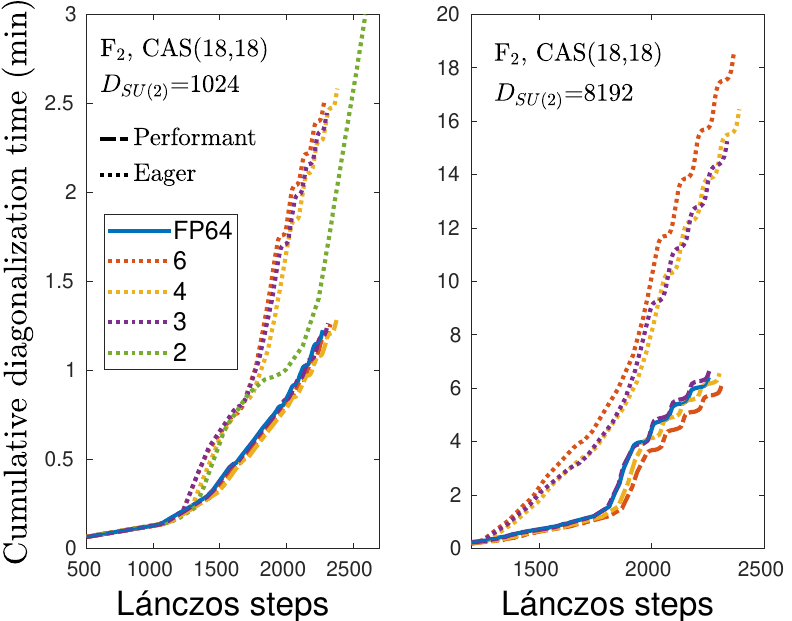}
    \caption{Cumulative diagonalization time in minutes as a function of L\'anczos steps 
    for simulations discussed in Fig.~\ref{fig:egs_f2_m1024} obtained on a DGX B200 system.
    In practice the performant mode (non-Eager mode), 
    where the system decides when it is faster to run emulation, is used.
    }
    \label{fig:time_f2_m1024}
\end{figure}

{\it Performance assessment:} 
In Fig.~\ref{fig:time_f2_m1024} we present 
the accumulated wall time in minutes for the diagonalization of the effective Hamiltonian for simulations discussed in Fig.~\ref{fig:egs_f2_m1024}.
It is evident that when eager mode is enforced, i.e., when almost all matrices employ emulation, the wall time increases significantly for $S\ge3$ slices compared to the native FP64 limit.
Without using eager mode, however, the FP64 profile is basically recovered and even a slightly lower wall time is found for $S\in\{4,6\}$ for the larger $D=8192$ simulations.
\begin{figure}
    \centering        \includegraphics[width=0.45\textwidth]{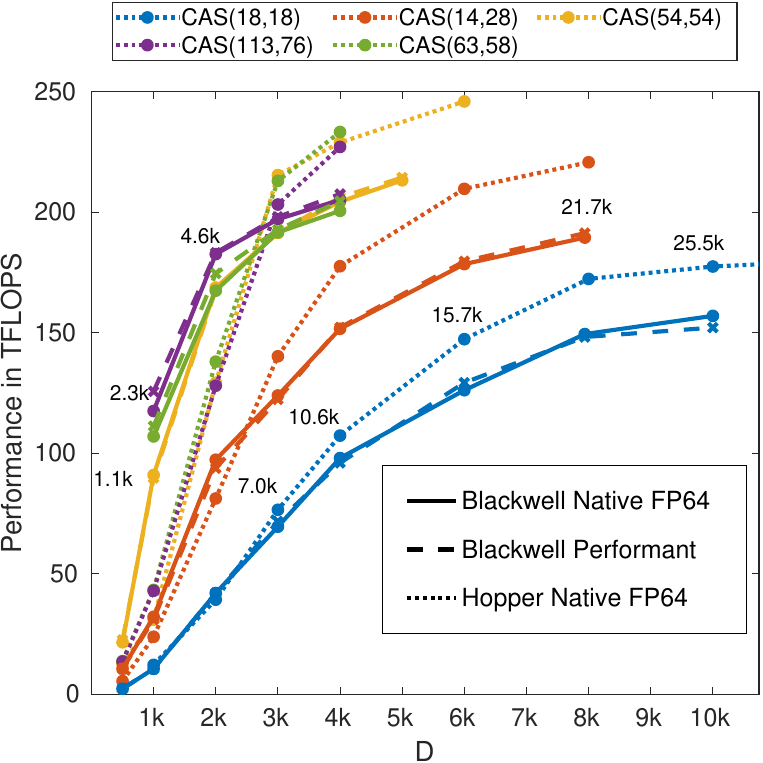}
    \caption{Benchmark results obtained via the SU(2) spin-adapted single node hybrid CPU plus multi-GPU DMRG calculations for the F$_2$ molecule on a CAS(18,18) orbital space~\cite{Legeza-2003a}, the
    N$_2$ molecule on a CAS(14,28) space~\cite{Chan-2004b}, FeMoco on CAS(54,54)~\cite{Reiher-2017} and CAS(113,76)~\cite{Li-2019} spaces, and P450 on CAS(63,58)~\cite{Goings-2022}.
    The solid lines correspond to calculations performed on a DGX B200 system via native FP64 precision, while dashed lines correspond to emulated performant mode.
    As a reference, the dotted lines trace the results obtained on a DGX H100 system~\cite{Menczer-2024b}.
    Numbers indicate the corresponding $U(1)$ bond dimension values, which are the same for the dotted, dashed, and the solid lines.}
     \label{fig:perf_b200}
\end{figure}
For completeness, we also present performance assessment in Fig.~\ref{fig:perf_b200} by comparing maximum performance measured in TFLOPS via the diagonalization of the effective Hamilton operator for the native FP64 limit on DGX H100 and DGX B200 systems.
For smaller $D$ values a significant increase in performance becomes apparent on the Blackwell node, while after a crossover, a 10-15\% decrease in performance is observed for larger $D$ values. 
On the DGX B200 system similar profiles have also been obtained via the emulated (performant) mode 
which for larger $D$ values and larger system sizes, $N$, 
slightly 
improve performance and efficiency. Therefore, we expect algorithms and implementations to improve and make a bigger impact on DMRG performance in the future.
A subsequent publication will include a more detailed performance analysis using an improved cuBLAS library and performant mode.

We remark that our hybrid CPU-multiGPU DMRG implementation can exploit almost the full power of the Blackwell system for large bond dimensions and system sizes, reflected by the fact that 90-95\% of the Thermal Design Power (TDP) was utilized, i.e. 900-950 Watts for each GPU card. Moreover, DGX B200 is very beneficial to fulfill the large memory demands of \textit{ab initio} DMRG, as it offers a total of 1.44 TB of GPU memory.

{\it Conclusion:} 
We presented numerical analysis and benchmark calculations of recent developments based on the Ozaki scheme for emulating FP64 arithmetic through the use of fixed-point compute resources by
employing the massively parallel spin-adapted \textit{ab initio} density matrix renormalization group (DMRG) method on selected strongly correlated chemical systems. 
By adjusting matrix ranks (bond dimensions) and system sizes, we performed a detailed error and performance analysis 
by approximating FP64 arithmetic with lower precision fixed-point elements
(referred to as “slices”)
and demonstrated that chemical accuracy can be reached via mixed-precision arithmetic using limited number of slices.  
However, for the half-like precision
limit (two slices)
the obtained ground state energy values can become unphysical and
fall below the exact reference energy, showing that such crude approximation is not acceptable.
Benchmark profiles obtained on the DGX B200 system showed only a slight decrease in performance rate compared to that of a DGX H100 node, which is expected 
to be compensated via the improved cuBLAS library and performant mode in the near future.
Finally, utilization of the presented mixed-precision arithmetic for orbital optimization via the GPU-aware DMRG-SCF framework ~\cite{Legeza-2025a} is straightforward.
Taken together, these highlight the efficient utilization of state-of-the-art Blackwell technology in tree-like
tensor network state electronic structure calculations, opening new research directions in materials sciences and beyond.
  
{\it Acknowledgments:} This work has been supported by the Hungarian National Research, Development and Innovation Office (NKFIH) through Grant Nos.~K134983 and TKP2021-NVA-04, and by the Center for Scalable and Predictive methods for Excitation and Correlated phenomena (SPEC), funded as part of the Computational Chemical Sciences, FWP 70942, by the U.S. Department of Energy (DOE), Office of Science, Office of Basic Energy Sciences, Division of Chemical Sciences, Geosciences, and Biosciences at Pacific Northwest National Laboratory.
\"O.L. acknowledges financial support
by the Hans Fischer Senior Fellowship programme funded by the Technical University
of Munich – Institute for Advanced Study.


%

\begin{mcitethebibliography}{113}
\bibitem[fp6()]{fp64-like}
In this paper we refer to the use of emulated DGEMM as emulated FP64, even when
  the number of bits used for emulation is fixed at a level that is
  insufficient for FP64 accuracy.\relax
\mciteBstWouldAddEndPunctfalse
\mciteSetBstMidEndSepPunct{\mcitedefaultmidpunct}
{}{\mcitedefaultseppunct}\relax
\EndOfBibitem
\end{mcitethebibliography}

\providecommand{\latin}[1]{#1}
\makeatletter
\providecommand{\doi}
  {\begingroup\let\do\@makeother\dospecials
  \catcode`\{=1 \catcode`\}=2 \doi@aux}
\providecommand{\doi@aux}[1]{\endgroup\texttt{#1}}
\makeatother
\providecommand*\mcitethebibliography{\thebibliography}
\csname @ifundefined\endcsname{endmcitethebibliography}
  {\let\endmcitethebibliography\endthebibliography}{}

\newpage

{\it Supporting Information}:
In the supporting information we present further numerical results supporting our numerical analysis and conclusions.

\begin{figure}[htb]
    \centering        \includegraphics[width=0.48\textwidth]{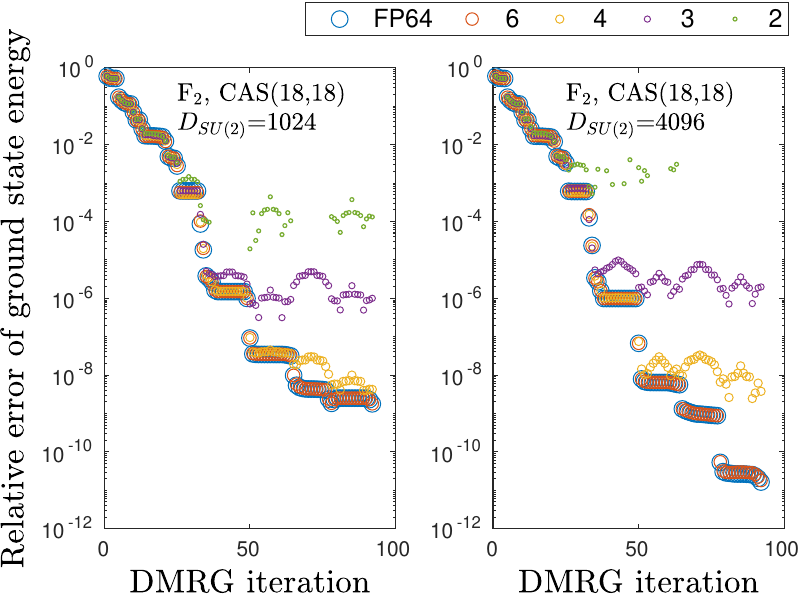}
    \caption{Similar to Fig.~\ref{fig:egs_f2_m1024} but obtained on a DGX H100 supercomputer.}   \label{fig:egs_f2_m1024_energy_h100}
\end{figure}

\begin{figure}[htb]
    \centering        \includegraphics[width=0.48\textwidth]{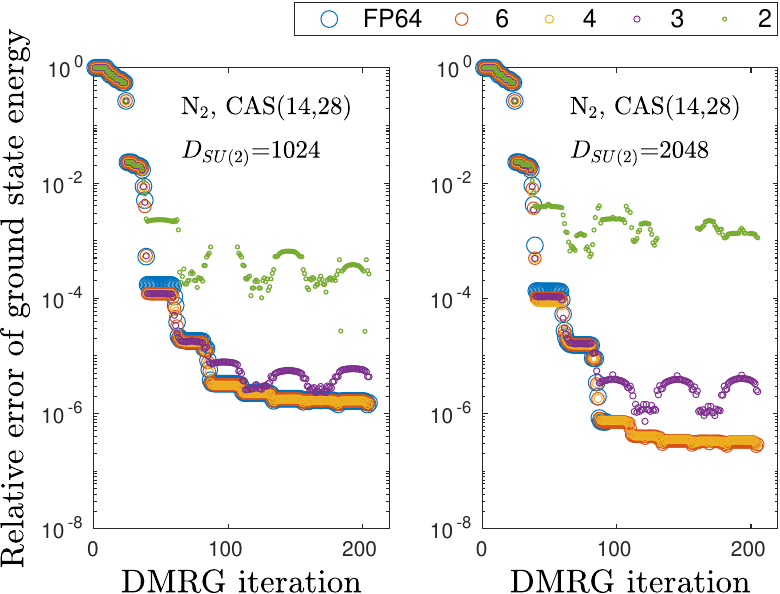}
    \caption{Similar to Fig.~\ref{fig:egs_n2_m1024} but obtained on a DGX H100 supercomputer.}   \label{fig:egs_n2_m1024_energy_h100}
\end{figure}

In Figs.~\ref{fig:egs_f2_m1024_energy_h100} and ~\ref{fig:egs_n2_m1024_energy_h100}  
we summarize results for systems discussed in Figs.~\ref{fig:egs_f2_m1024} and
~\ref{fig:egs_n2_m1024} but 
obtained on a DGX H100 supercomputer.

In Fig.~\ref{fig:egs_f2_m1024_ren_svd_s3} the relative error of the ground state energy, $\Delta E_{\rm rel}$, as a function of DMRG iteration steps is shown for the F$_2$ molecule in a CAS(18,18) model space using bond dimension $D_{SU(2)}=1024$, $S=3$ slices, setting the residual error threshold to $\varepsilon=10^{-5}$ in the L\'anczos diagonalization
and by switching between CPU and GPU implementations for the renormalization (network contraction) and SVD algorithmic parts.
We observe that
employing a CPU or GPU based renormalization procedure, which is mainly based on DGEMM operations, has only minor effect on the obtained convergence profile. 
In contrast to this, switching from the NVIDIA reduced precision cuSOLVER-based GPU implementation of the SVD step (diagonalization of the reduced density matrix) to the CPU variant the accuracy improves significantly.
Therefore,
the error accumulated via SVD using small number of slices, $S=3$, determines the overall convergence. 
Note that similar improvement is achieved by using $S\in\{7,8\}$ slices or native FP64 in cuSOLVER.

In Fig.~\ref{fig:egs_f2_m1024_lanczos} the relative error of the ground state energy, $\Delta E_{\rm rel}$, as a function of DMRG iteration steps is shown for the F$_2$ molecule in a CAS(18,18) model space 
using $D_{SU(2)}=1024$ and $S=2$ slices
for various residual error threshold values, $\varepsilon$, employed in the L\'anczos method.
By reducing the residual error,
$\varepsilon$, to $10^{-4},\ldots,10^{-2}$ the number of non-variational eigenvalues disappeared, leading to an oscillating curve in relative energy in the range of $10^{-4}$ as in Fig.~\ref{fig:egs_f2_m1024} but without ``missing" data points. 

In Fig.~\ref{fig:egs_f2_m1024_ren_svd} we present similar analysis as shown in Fig.~\ref{fig:egs_f2_m1024_ren_svd_s3} but for $S=2$ slices.
Similarly, as discussed for Fig.~\ref{fig:egs_f2_m1024_ren_svd_s3}, employing a CPU or GPU based renormalization procedure
has only minor effect on the obtained convergence profile. 
In contrast to this, switching from the reduced precision cuSOLVER-based GPU implementation of the SVD step (diagonalization of the reduced density matrix) to the CPU variant the accuracy improves significantly. Therefore, as expected, the large error accumulated via SVD using only $S=2$ slices prohibits DMRG to converge. 

\end{document}